\documentclass[10pt,aps,twocolumn,prd,superscriptaddress,noshowpacs,nofootinbib,noshowkeys,floatfix]{revtex4}
\usepackage[dvips]{graphics,graphicx}
\usepackage[colorlinks=true,linktocpage=true,linkcolor=blue,citecolor=blue]{hyperref}
\usepackage[usenames,dvipsnames]{color}
\usepackage{amsmath, amssymb}
\usepackage{multirow}
\usepackage{longtable}
\usepackage{color}
\usepackage[normalem]{ulem}  % \sout{old text} for strikeout

\renewcommand\sout{\bgroup \color{blue} \ULdepth=-.5ex \ULset}
\begin{document}

\title{Finite volume corrections and low momentum cuts in the thermodynamics of quantum gases }
%\date{\today}
\author{Krzysztof Redlich}
\affiliation{
Institute of Theoretical Physics, University of Wroc\l aw\\ PL-50-204 Wroc\l aw, Poland}
\author{Kacper  Zalewski}
\affiliation{Institute of Nuclear Physics, Polish Academy of Sciences\\ PL-31-342 Krak\'ow, Poland}
\affiliation{
   Institute of Physics,  Jagellonian University \\
     PL-30-059 Krak\'ow,  Poland
   }
%\maketitle
\begin{abstract}
{The conjecture, that  the finite  volume corrections  to the thermodynamic functions  can be correctly reproduced by using the thermodynamic limit
with low particle momenta cutoff is examined in a very transparent example of an ideal boson gas in one dimension.}
 We show that this  {conjecture} is always true in principle,  { and derive convenient relations   for the momentum  cutoff dependence on thermal  parameters in the asymptotic limits of large and small  volume.}
\end{abstract}
\newpage

\maketitle
\section{Introduction}

{In the phenomenological  analysis  of particle production in high energy heavy ion collisions it was shown that the thermal statistical models provide a very satisfactory description of particle yields measured in the experiments \cite{anton1,anton2,anton3,anton4,anton5}. The statistical operator in these models  is usually constructed  as that for a  mixture of ideal gases constrained by the conservation laws.
The success of such statistical  models has recently got a theoretical support from lattice QCD (LQCD), showing that they are also capable  to  quantify
 the equation of state and
 different fluctuation observables  in the hadronic phase obtained in LQCD \cite{karsch1,Bazavov,karsch,lgt1}.

In heavy ion  collisions, however,  we are dealing  with a finite system, thus in the thermal analysis of data,   the effects related with the finite volume should be accounted for. Indeed, its was shown,  that such corrections are crucial in small systems,  and they have been quantified in the context of exact conservation laws \cite{anton4,anton5,can1}. Recently, the influence of the finite volume  effects were also addressed in the context of fluctuation  observables in the momentum space \cite{BRS,KMR}. There,  is was conjectured, that  the final volume corrections  in the thermodynamic observables can be correctly reproduced by using the thermodynamic limit
with low particle momenta cutoff.

In the statistical thermodynamics of gases, in the thermodynamic limit,  where the volume  and the
 number of particles  tend to infinity at fixed particle density and temperature, the
 thermodynamic functions are represented as integral over momentum space. Following numerical results of   Refs.  \cite{BRS},   \cite{KMR} and \cite{EKS}, one expects that the    finite volume corrections for the
 thermodynamics of quantum gases can be approximated by cutting off the low momentum regions in
 these integrals formulae.

  In the present paper we use a very  transparent  model of   one-dimensional gas
 of noninteracting bosons, to discuss this conjecture. We find that for a significant range
 of volumes the approximation is valid  with a simple intuitively plausible momentum cutoff. For smaller
 volumes,  the finite volume effects can still  be reproduced by introducing  a momentum cutoff, but
 the position of the cutoff becomes a complicated function of the parameters of the system. For sufficiently large volume, the momentum cutoff $k_c=\pi/2L$  is found to be simply connected with the  system size $L$,  which makes it
 convenient for
  the phenomenological application of statistical models to  the  finite systems.}

 \section{Thermodynamic functions at finite volume}

Let us consider an ideal gas of identical bosons  {of mass $m$} enclosed in a one-dimensional box of length $L$.
Assuming  the boundary conditions for the single-particle
 wave function,

\begin{equation}\label{}
  \psi(0) = \psi(L) = 0,
\end{equation}
one finds,  that each single-particle state is unambiguously defined by its energy,

\begin{equation}\label{}
  E_n = \sqrt{\left(\frac{\pi n}{L}\right)^2 + m^2};\qquad n = 1,2,\ldots.
\end{equation}

All the thermodynamic functions of the gas can be obtained from the  {thermodynamic} potential

\begin{equation}\label{eq3}
\Omega(T,V,\mu) =  T\sum_{n=1}^\infty \log\left(1 - e^{-\beta (E_n - \mu)}\right),
\end{equation}
where, as usual, $\beta = T^{-1}$. It is convenient to introduce the function

\begin{equation}\label{}
\Omega_n(T,V,N) = T\log\left(1 - e^{-\beta (E_n - \mu)}\right),
\end{equation}
where $n$ is a continuous variable ranging from zero to infinity.

For $L$ tending to infinity, the  function $\Omega(T,V,\mu)$ goes over into the integral
\begin{equation}\label{eq5}
  \Omega_{int}(T,V,\mu) = T\int_0^{\infty}\;\Omega_n(T,V,\mu)\;dn.
\end{equation}
Making the substitution,
%\begin{equation}\label{}
 $ z = \frac{\beta\pi}{L}n$,
%\end{equation}
 {the potential}  can be {also}  {expressed}, as
\begin{equation}\label{}
  \Omega_{int}(T,V,\mu) = \frac{L}{\beta\pi}\int_0^{\infty}\;\log\left(1 - e^{-\left(E(z) -
  \beta\mu\right)}\right)\;dz,
\end{equation}
where,
%\begin{equation}\label{}
$  E(z) = \sqrt{z^2 + \beta^2m^2}.$
%\end{equation}

\begin{figure}[!t]
\centering
\includegraphics[width=2.89in,height=2.905in,angle=-90]{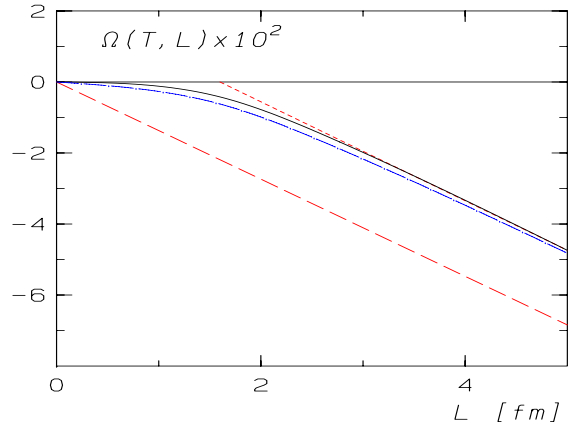}
\vskip -0.9cm
\caption{Comparison of the thermodynamic potential  $\Omega(120,L,0)$ (full line) with its approximations:  $\Omega_{int}(120,L,0)$ (long-dashed  line), $\Omega_{cor}(120,L,0)$ (short-dashed line) and $\Omega_{cut}(120,L,0,0.5)$ (dashed-dotted  line), {\it see text}. }
\label{figoic}
\end{figure}

Since $\Omega_{int}$ is proportional to the volume $L$, the thermodynamic identity

\begin{equation}\label{pfromo}
  p = -\left(\frac{\partial \Omega}{\partial L}\right)
\end{equation}
yields, in the integral approximation, a well-known relation,
%\marginpar{\emph{ploint}}
\begin{equation}\label{ploint}
  p_{int} =-{1\over L} \Omega_{int},
\end{equation}
{{for the thermodynamic pressure in one-dimension}.

\subsection{ The approximations of the final volume  potential}

In the following,  we focus on  the finite volume corrections to thermodynamics of an ideal boson gas
and discuss their  relation to the particle momentum cutoff.
We examine  different approximations of the  potential (\ref{eq3}) and consider them  to be {\it good} if, for a given set of parameters,  they differ by less than one percent from the exact result.
For   numerical calculations we take { the particle mass $m=140~\mbox{MeV}$, the temperature $T = 120 ~\mbox{MeV}$, and assume vanishing chemical potential $\mu = 0$.}
%\marginpar{\emph{parame}}\begin{equation}\label{parame}
%\end{equation}

{In Fig. \ref{figoic} we show the  $L$-dependence of $\Omega(120,L,0)$ and $\Omega_{int}(120,L,0)$ from Eqs. (\ref{eq3}) and (\ref{eq5}), respectively.}
 The approximation of the  sum   (\ref{eq3}) by the integral (\ref{eq5})  is valid at very large $L$, and it is {\it good} only  at   $ L > 166~\mbox{fm}$.}
{However,  one  finds that} an approximation to $\Omega$,
% \marginpar{\emph{aprtra}}
 \begin{equation}\label{aprtra}
  \Omega_{cor} = \Omega_{int} - \frac{1}{2}\Omega_0.
\end{equation}
 is much better than $\Omega_{int}$.
 This may be interpreted as an application of the Euler-Maclaurin series for the difference
 between the sum and the corresponding integral. Only one term in this series is different from
 zero. Thus, the series is convergent, but unfortunately not to the expected limit. The same
 formula can be  obtained by applying the well-known method of trapezes to approximate the
 integral.

 The correction term  in Eq. (\ref{aprtra}),  is calculated from Eq. (\ref{eq3}),  and  at $T=120~\mbox{MeV}$, one gets
% \begin{equation}\label{}
$\Omega_0 = -44.8~\mbox{MeV}.$
%\end{equation}
As shown in Fig. \ref{figoic},  the improvement  {of}  the approximation of $\Omega$ by $\Omega_{cor}$ is {very significant, and
is  much
 better   than $\Omega_{int}$,   since   it
is   {\it good}}  already for
%\begin{equation}\label{}
 $ L > 3.7~\mbox{fm}.$
%\end{equation}

\subsubsection{Momentum cutoff and the finite volume effects}
{The thermodynamic potential of an ideal boson gas enclosed in a finite volume is a  transparent example of a system
where the conjecture,  that by cutting
 off the low momenta
in the thermodynamic limit one reproduces the exact results for the
 bounded system,  can be verified.
In general, we consider}
 the problem of finding a cutoff $n_c$ such that
%\marginpar{\emph{omecut}}
\begin{eqnarray}\label{omecut}
\Omega_{cut}(T,V,\mu,n_c) &\equiv&\nonumber  \\
T\int_{n_c}^{\infty}\;\Omega_n(T,V,\mu)\;dn = \Omega(T,V,\mu).
\end{eqnarray}
{The} function $\Omega_n$ is a continuous, monotonically increasing function of $\frac{n}{L}$. Therefore,
\begin{eqnarray}\label{}
  \int_n^{n-1}\;\Omega_m(T,L,\mu)\;dm &<& \Omega_n(T,L,\mu)< \nonumber\\
   \int_n^{n+1}\;\Omega_m(T,L,\mu)\;dm,
\end{eqnarray}
and consequently

\begin{equation}\label{}
  \Omega_{cut}(T,L,\mu,0) < \Omega(T,L,\mu) < \Omega_{cut}(T,L,\mu,1).
\end{equation}
This implies that, at {any} given $T$, $\mu$  and $L$, it is always possible to find, in the range
\begin{equation}\label{}
  0 < n_c < 1,
\end{equation}
a value of $n_c$  satisfying  the   relation (\ref{omecut}).

We  {consider first}  a large-$L$ limit, in which the difference between $\Omega_{int}$
and $\Omega_{cut}$ can be approximated by a linear function of $n_c$. Assuming that in this
region    $\Omega_{cor}$ is a {\it good} approximation to $\Omega$, one gets the  intuitively expected result (see e.g. \cite{BRS})
\begin{equation}\label{encut}
  n_c =\frac{1}{2}.
\end{equation}
 The particle momentum
%\begin{equation}\label{}
 $ k_n = \frac{\pi}{L}n$,
%\end{equation}
{thus} the cutoff at $n_c$ is equivalent to the  cutoff in the particle momentum at the corresponding
\begin{equation}\label{}
k_c=\frac{\pi}{2L},
\end{equation}
 which is consistent with the numerical value  found  in Ref.  \cite{KMR} for the transverse momentum cutoff in three dimensions.

A comparison of $\Omega_{cut}$ for this cutoff with the exact result, and with the two other approximations,   {discussed in the previous section},  is shown in Fig. \ref{figoic}.
  The approximation
{of   $\Omega(120,L,0)$  by $\Omega_{cut}(120,L,0,\frac{1}{2})$}
is {\it good} for
%\begin{equation}\label{}
 $ L > 5.6~\mbox{fm}.$
%\end{equation}
Actually, the approximation to the correction term $\Omega_0$ is {\it good} only for $L >
 8.4$ fm, but with increasing $L$ the relative importance of the correction term decreases.

Another {transparent expression } for $n_c$ can be obtained  {in the $L\to 0$ limit, where}  the ratio $\frac{L}{n_c}$ is so small that the approximation,

\begin{equation}\label{conlnc}
  \log\left(1 - e^{-\beta(E_n - \mu)}\right) \approx -e^{\beta\left(\frac{n\pi}{L}-\mu\right)}
\end{equation}
{is valid.}  For the values  of the parameters  used in the previous section, this is the case when
%\marginpar{\emph{conlnc}}\begin{equation}\label{conlnc}
 $ L < 0.076 n_c ~\mbox{fm}.$
%\end{equation}
Then, the   condition (\ref{omecut}) for $n_c$ can be approximated by

\begin{equation}\label{}
  \int_{n_c}^\infty\;e^{-\frac{\beta\pi n}{L}}\;dn = e^{-\frac{\beta\pi}{L}}
\end{equation}
and yields

\begin{equation}\label{low}
  n_c = 1 + \frac{L}{\beta\pi}\log\frac{L}{\beta}{\pi}.
\end{equation}
The requirement of consistency with the condition (\ref{conlnc}) gives the limitation
%\begin{equation}\label{}
 $ L < 0.076~\mbox{fm}.$
%\end{equation}
{Thus,} this low-$L$ formula (\ref{low})  is of little interest for direct physical applications.

%%%%%%%%%%%%%%%%%%%%%%%%%%%

The {numerical} results for    $n_c$ as a function of $L$ are  shown in Fig. 2 %\ref{figncl}.
 It is seen,  that $n_c$ decreases monotonically from {unity}  at $L \rightarrow 0$ to its asymptotic value $n_c=\frac{1}{2}$ at large $L$.
%From Fig. \ref{figoic}  it is clear that this value of $n_c$ is {\it good}  for  $L>5.6$ fm.
 For higher temperatures, the validity limit of this approximation is shifted to lower values of $L$. 
%This is because, the cutoff  $n_c$ is only  a function of dimensional parameter $LT$.

The results for the number of particles, energy and entropy are qualitatively similar to the
results for the potential $\Omega$. They can be obtained either by studying the exact expressions
for these quantities, or by using the results for the potential $\Omega$ and suitable
thermodynamic identities.

 Note  that  some identities derived using specific assumptions about the
system, e.g. $\Omega = -pL$, cannot be used blindly.
For the pressure,  the situation is somewhat different. Using Eq. (\ref{pfromo}) one  gets
\begin{equation}\label{}
   p(T,L,\mu)= \frac{\pi^2}{L^3}\sum_{n=1}^\infty\;\frac{n^2}
   {\left(e^{\beta(E_n-\mu)}-1\right)E_n}.
\end{equation}
Defining the function of $n$,

\begin{equation}\label{}
  p_n(T,L,\mu) = \frac{\pi^2}{L^3}\frac{n^2}{\left(e^{\beta(E_n-\mu)}-1\right)E_n}
\end{equation}
we see that
%\begin{equation}\label{}
 $ p_0(T,L,\mu) = 0$,
%\end{equation}
  {thus} no subtraction should be made. An alternative derivation of this result follows from the
remark, valid in the region where $\Omega_{cor}$ is a {\it good} approximation to $\Omega$, that the
difference between $\Omega_{int}$ and $\Omega$ does not depend on $L$, and thus its derivative
with respect to $L$ does not contribute to the pressure. Consequently, the integral approximation

\begin{eqnarray}\label{}
  p_{int}(T,L,\mu) &=& \int_0^\infty\;p_n(T,L,\mu)\;dn  =\nonumber \\ &&\frac{1}{\pi}\int_0^\infty\;\frac{z^2dz}{
  \left(e^{\left(E(z)-\beta\mu\right)}-1\right)E(z)}
\end{eqnarray}
is much better than in the previous case { of the thermodynamic potential}.

\begin{figure*}[ht]\label{figncl}
 \centering
 {\vskip -0.8cm {\includegraphics[width=3.21in,angle=-90]{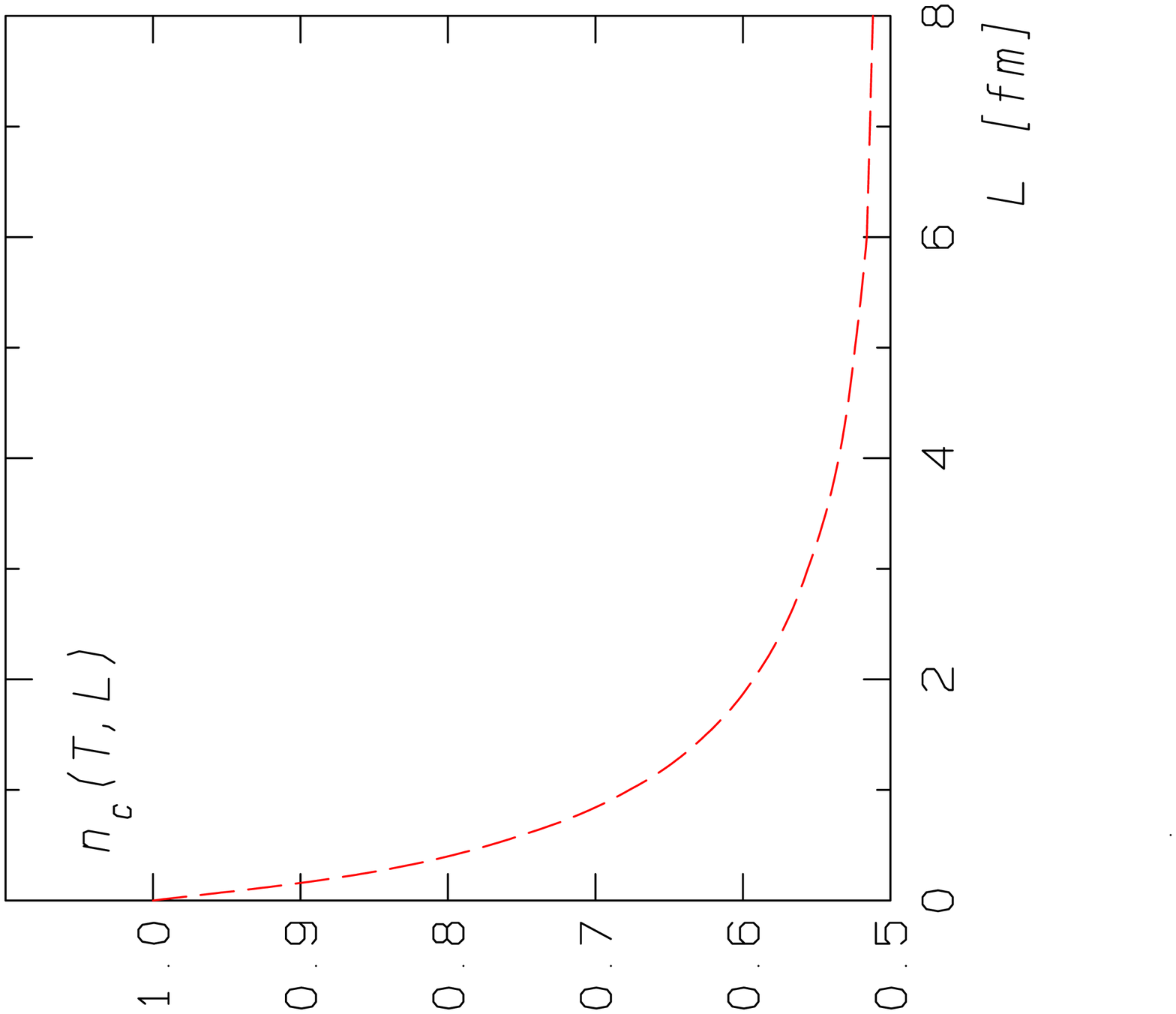}\hskip 2.7cm}}
{ \includegraphics[width=2.95in,angle=-90]{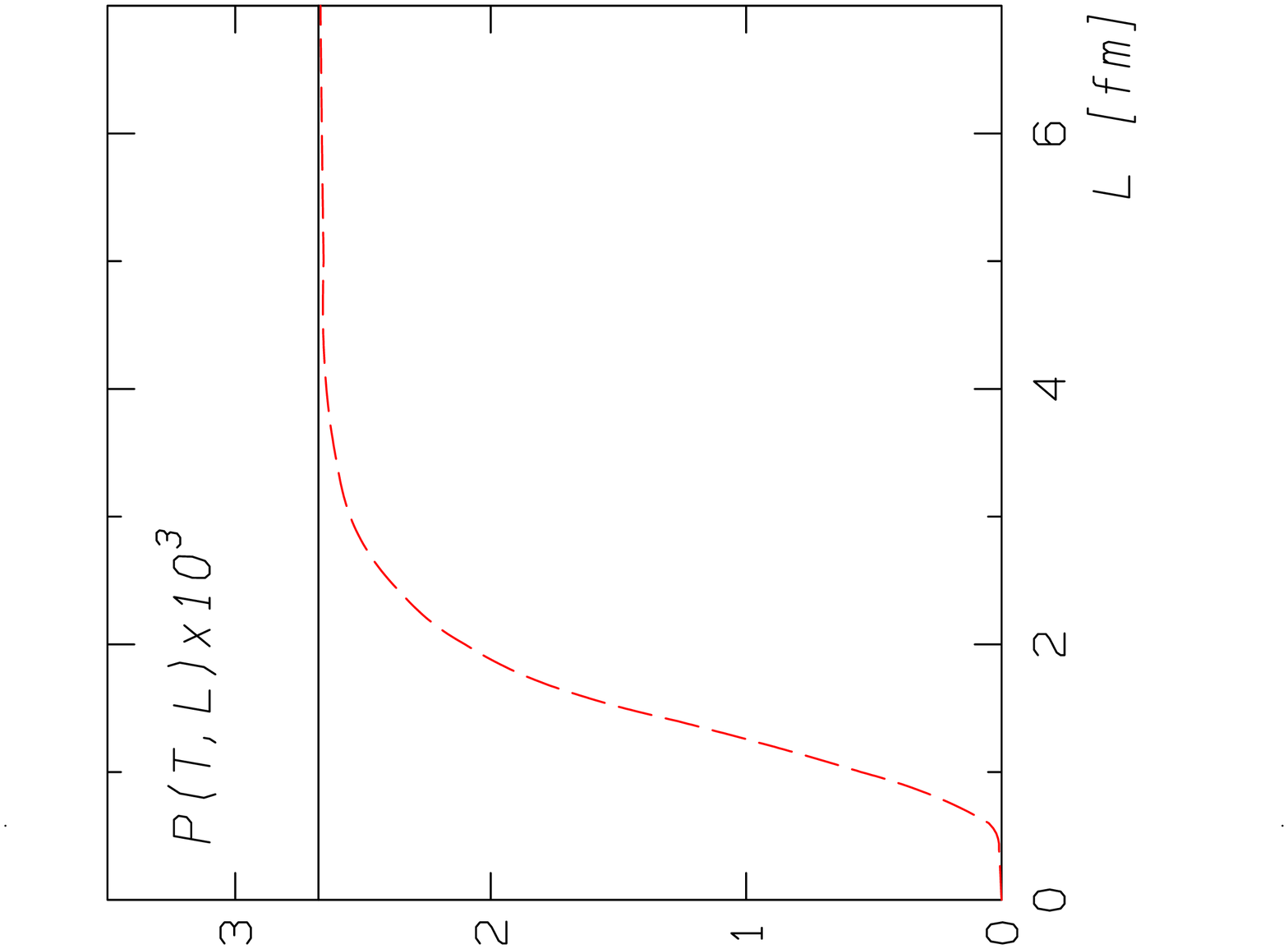}}
 \vskip -1.3cm
 \caption{Left-hand figure: The $L$-dependence of the cutoff parameter $n_c$.
 Right-hand figure:
 Comparison of the function $p(120,L,0)$ (dashed  line) with the approximation $p_{int}(120,L,0)$ (full line).
}
 \end{figure*}

The $L$ dependence of $p_{int}$ and $p$ is compared in
 Fig. 2.
 %\ref{figncl}.
   The integral approximation is {\it good} in the whole region {where}
%\begin{equation}\label{}
 $ L > 4.4 ~{\rm fm.}$
%\end{equation}
Note,  that  since $\Omega_{cor}$ is a good approximation to $\Omega$ in this region, we get from
Eq.  (\ref{ploint}) the relation

\begin{equation}\label{}
  \Omega + \frac{1}{2}\Omega_0 \approx -pL.
\end{equation}

{The numerical results for the cutoff parameter and the limits of the applicability of different approximations of the  finite  volume thermodynamic functions where calculated  at fixed temperature. However, the temperature dependence of the momentum cutoff function is  straightforward. }

\section{Conclusions}
{For  an ideal boson gas in one-dimension,
we have demonstrated  that the finite volume corrections
 to thermodynamic functions
 can be exactly reproduced in the thermodynamical limit by making suitable low-momentum cuts in the particle phase-space.

 The calculations with the cutoff parameter  are greatly simplified when {it} does not depend on  the  model {parameters}. We have shown, that for the one-dimensional model considered here,  this  is the case  when the length of the vessel  $L$ is no less than a
few fermis.
There, the momentum cutoff $k_c=\pi/2L$  is simply related with the volume,  which makes it very transparent  in the  numerical analysis of finite volume effects in  different observables. For smaller
 volumes,  the finite volume effects can still  be reproduced by introducing  a momentum cutoff, but
 the position of the cutoff becomes a more complicated function of the parameters of the system. Nevertheless, in the limit  of $L\to 0$, a compact analytic expression of the momentum cutoff has been   derived.

The above results support the conjecture     that in the application of statistical models to particle production in heavy ion collisions the importance of the finite volume effects can be studied in the thermodynamic limit by implementing the cutoff in the momentum phase space of  particles. From the discussion  presented here, it is transparent, that the finite volume corrections  are relevant for small systems when their size is less than a few fermis. {This is the case in a medium crated in elementary,  or   peripheral  nucleus-nucleus collisions.  }}

\section*{ Acknowledgments}
One of the authors (KZ) was partly supported by the Polish National Science Center (NCN), under  grant DEC-2013/09/B/ST2/00497. K.R. acknowledges  support of  the Polish Science Center (NCN) under Maestro grant DEC-2013/10/A/ST2/00106.


\begin{thebibliography}{99}
 \bibitem{anton1}
 A. Andronic, Int. J. Mod. Phys. A
{\bf  29}, 1430047
(2014).

\bibitem{anton2}
 M. Floris, Nucl. Phys. A
{\bf 931},  103
(2014).
 \bibitem{anton3}
   A.~Andronic, P.~Braun-Munzinger and J.~Stachel,
  %``Thermal hadron production in relativistic nuclear collisions: The Hadron mass spectrum, the horn, and the QCD phase transition,''
  Phys.\ Lett.\ B {\bf 673}, 142 (2009).


\bibitem{anton4}
P.  Braun-Munzinger,  K.  Redlich,  and  J.  Stachel,  Particle  production  in  heavy  ion  collisions,  in:  R.  C.  Hwa
and X.-N. Wang (Eds.), Quark-Gluon Plasma 3, World
Scienti c,  Singapore,  2004,  pp.  491-599.  e-Print:   nucl-
th/0304013.




\bibitem{anton5}
 F.~Becattini, P.~Castorina, A.~Milov and H.~Satz,
  %``A Comparative analysis of statistical hadron production,''
  Eur.\ Phys.\ J.\ C {\bf 66}, 377 (2010).


\bibitem{karsch1}
	 F.~Karsch,
	 %``Moments of charge fluctuations, pseudo-critical temperatures and freeze-out in heavy ion collisions,''
	 J.\ Phys.\ G {\bf 38}, 124098 (2011).

\bibitem{Bazavov}
  A.~Bazavov {\it et al.} [HotQCD Collaboration],
  %``Fluctuations and Correlations of net baryon number, electric charge, and strangeness: A comparison of lattice QCD results with the hadron resonance gas model,''
  Phys.\ Rev.\ D {\bf 86}, 034509 (2012).
  A.~Bazavov {\it et al.} [HotQCD Collaboration],
  %``Equation of state in ( 2+1 )-flavor QCD,''
  Phys.\ Rev.\ D {\bf 90}, 094503 (2014).
\bibitem{karsch}
  F.~Karsch,
  %``Thermodynamics of strong interaction matter from lattice QCD and the hadron resonance gas model,''
  Acta Phys.\ Polon.\ Supp.\  {\bf 7}, no. 1, 117 (2014).

   \bibitem{lgt1}
	 S.~Borsanyi, Z.~Fodor, S.~D.~Katz, S.~Krieg, C.~Ratti and K.~K.~Szabo,
	 %``Freeze-out parameters: lattice meets experiment,''
	 Phys.\ Rev.\ Lett.\  {\bf 111}, 062005 (2013).


\bibitem{can1}
   S.~Hamieh, K.~Redlich and A.~Tounsi,
  %``Canonical description of strangeness enhancement from p-A to Pb Pb collisions,''
  Phys.\ Lett.\ B {\bf 486}, 61 (2000).


\bibitem{BRS}
 A.~Bhattacharyya, R.~Ray, S.~Samanta and S.~Sur,
  %``Thermodynamics and fluctuations of conserved charges in a hadron resonance gas model in a finite volume,''
  Phys.\ Rev.\ C {\bf 91}, no. 4, 041901 (2015).

\bibitem{KMR}
F.~Karsch, K.~Morita and K.~Redlich,
  %``Effects of kinematic cuts on net-electric charge fluctuations,''
  Phys.\ Rev.\ C {\bf 93}, no. 3, 034907 (2016).
  \bibitem{EKS}
J.~Engels, F.~Karsch and H.~Satz,
  %``Finite Size Effects in Euclidean Lattice Thermodynamics for Noninteracting Bose and Fermi Systems,''
  Nucl.\ Phys.\ B {\bf 205}, 239 (1982).
\end{thebibliography}
\end{document}